# Modelling Electricity Consumption in Office Buildings: An Agent Based Approach

Tao Zhang, Peer-Olaf Siebers, Uwe Aickelin
Intelligent Modelling & Analysis Group
School of Computer Science, University of Nottingham
Tao.Zhang@nottingham.ac.uk

## Abstract

In this paper, we develop an agent-based model which integrates four important elements, i.e. organisational energy management policies/regulations, energy management technologies, electric appliances and equipment, and human behaviour, to simulate the electricity consumption in office buildings. Based on a case study, we use this model to test the effectiveness of different electricity management strategies, and solve practical office electricity consumption problems. This paper theoretically contributes to an integration of the four elements involved in the complex organisational issue of office electricity consumption, and practically contributes to an application of an agent-based approach for office building electricity consumption study.

## 1. Introduction

In the UK and many other industrialised countries, offices, as a basic unit for work buildings, are intensively distributed in big cities and urban areas. As climate change becomes a very important global issue, the UK government has set a target of cutting CO2 emission by 34% of 1990 levels by 2020. In the UK, the energy consumed in the service sector took up 14% of overall energy consumption of the whole country in 2001. Most of the energy for the service sector is used in various kinds of offices for heating, lighting, computing, catering and hot water. Thus, energy consumption in office buildings is one of the research areas which have significant importance for meeting the UK government's 2020 target.

Practically, energy consumption in a modern office building is a very complex organisational issue involving four important elements:

- Energy management policies/regulations made by the energy management division of an organisation

- Energy management technologies installed in the office building (e.g. metering, monitoring, and automation of switch-on/off technologies)

- Types and numbers of the electric equipment and appliances in the office building (e.g. lights, computers and heaters)

- Energy users' behaviour of using electric equipment and appliances in the office building.

The four elements interact (Figure 1) in the following way: The energy management division makes energy management policies/regulations based on the energy management technologies installed in the building; energy management technologies monitor and control the energy consumed by electric equipment and appliances, and also influence the behaviour of energy users in the building; energy users' behaviour of using electric equipment and appliances directly cause energy consumption.

Yet in the UK the energy consumption in office buildings has been primarily administered by the Energy Performance of Buildings Directive (EPBD), in which the National Calculation Method (NCM) is defined. The NCM is a procedure for demonstrating compliance with the Building Regulations for buildings other than dwellings. The NCM proposes that by calculating the annual energy use for a proposed building and comparing it with the energy use of a comparable "notional" building, an "asset rating" can be produced in accordance with the EPBD. The calculations in the NCM make use of standard sets of data for different activity areas and call on common databases of construction and service elements. The calculations are carried out by approved simulation software packages (e.g. Operational Rating Calculation (ORCalc), Front-end Interface for the Simplified Building Energy Model engine (FI-SBEM), and Dynamic Simulation Modelling (DSM)).

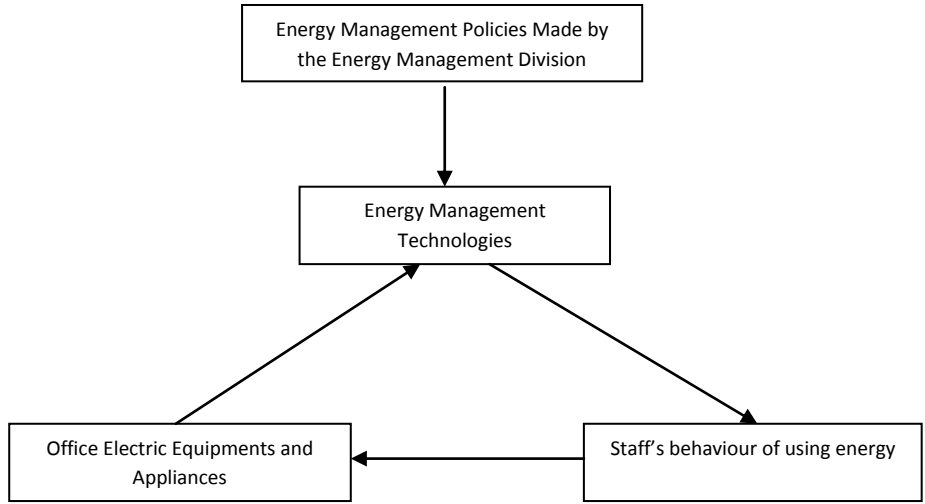

Figure 1: The Four Elements in Office Energy Consumption

Although the NCM is a powerful tool for office building management, it does not consider the areas of organisational energy management policies/regulations and human factors (i.e. energy users' behaviour) which are very important elements influencing office building energy consumption. Looking at the literature in building energy research, there has been some interest in the behavioural aspects of energy use [e.g. 1, 2, 3, 4]. Generally these studies are empirical models based on measurements in practice. Hoes et al. [5] evaluate the effect of user behaviour on building energy performance and propose a decision methodology to optimise the design of buildings which have a known close interaction with its users. The study is a numerical simulation that tends to provide a guideline for building design. It does not reveal the dynamic processes of energy user behaviour and their interactions, and how these dynamic processes contribute to the overall energy consumption of an office building. Motivated by a desire to comprehensively understand the complex organisational issue of office energy consumption, the **first objective** of the research reported in this paper is to provide a dynamic computational simulation model which integrates the aforementioned four interactive elements involved in office energy consumption.

Each organisation faces a dilemma in terms of energy consumption. On the one hand, it has to consume energy to satisfactorily meet the energy needs of staff and maintain comfort standards in its office buildings. One the other hand, it has to minimise its energy consumption through effective organisational energy management policies/regulations, in order to reduce energy bills. This dilemma presents a discord between energy users' behaviour and organisational energy management policies. To solve this discord, an organisation needs to make a balanced energy management decision between the two. Although the methodology proposed by Hoes et al [5] provides a guideline for new building design, in terms of aiding the energy management policy/strategy making for existing human-centric buildings (i.e. buildings which have large number of users) the methodology appears to be less robust. In recent years, agent-based simulation as a powerful decision support tool has been brought into the attention of researchers in building energy research areas [e.g. 6, 7]. Drawing on the idea of agents, a **second objective** of the research reported in this paper is to develop a multi-agent decision-making framework to help organisations make proper energy management policies/regulations to deal with this dilemma: maintaining efficient energy consumption to satisfactorily meet staff members' energy needs, whilst minimizing the energy consumption within the whole organisation, without significant investments on or changes to the current energy management technologies.

Practically, energy consumption in a modern office building has two parts: gas consumption, which does not exist in some types of office buildings; and electricity consumption, which generally exists in almost all types of office buildings. In this paper, we specifically focus on studying electricity consumption in office buildings. The paper is structured as follows. In the second section, we briefly introduce agent-based simulation method and the rationale of using agent-based modelling in studying office building electricity consumption. In the third section, we describe our agent-based model of office building electricity consumption based on a case study in an academic building in Jubilee Campus, University of Nottingham, and present the questions that we are going to study with the model. In the fourth section, we analyse the outputs from the simulation, and draw some electricity management strategy implications. In the fifth section, we discuss the model, and in the sixth section, we summarise and conclude the study.

## 2. Agent-Based Simulation: Methodology and Modelling Rationale

### 2.1 Agent-Based Simulation

In complexity science, agents are the constituent units of a complex adaptive system (CAS); agents are autonomous, intelligently behave on their own, and interact with each other in a CAS [8]. The intelligent behaviour and interactions of agents produce the global behaviour of a CAS. However, this type of global behaviour of a CAS cannot be conversely traced back to the behaviour and interactions of its constituent agents. Agents can be software programmes, machines, human beings, societies, or anything that is capable of intelligent behaviour [8]. Agent-based simulation is a computational modelling approach to study CASs. An agent-based model is composed of individual agents, commonly implemented in software as objects. Agent objects have states and rules of behaviour. Running an agent-based model simply amounts to instantiating an agent population, letting the agents behave and interact, and observing what happens globally [9]. Thus a unique advantage of agent-based simulation is that almost all behavioural attributes of agents can be captured and modelled. Agent-based simulation is widely adopted in studying CASs, particularly those with intelligent human beings (e.g. markets, societies, and organisations; for further information about agent-based simulation and its applications, please see [10, 11])

### 2.2 Modelling Rationale

The electricity consumption in an office building is caused by the operation of various types of electric equipment and appliances (e.g. electric heaters, computing equipment and lights) in the office, which in turn are controlled by the electricity users' behaviour. The electricity users interact, and their interactions can influence their behaviour of using electric appliances. The electricity users, electric appliances and the office building environment constitute a CAS which is well suited to agent-based modelling.

Firth et al [12] carry out a study on the types of home electric appliances that people use most frequently and how these electric appliances contribute to the overall electricity consumption in a household. They classified home electric appliances into four categories based on their pattern of use:

- Continuous appliances: Refer to electrical appliances such as clocks, burglar alarms and broadband Internet modems which require a constant amount of electricity.

- Standby appliances: Refer to electrical appliances such as televisions and game consoles which have three modes of operation: in use, on standby, or switched off; standby appliances consume electricity when they are in the modes of "in use" and "on standby", and some time even in the mode of "switched off" (e.g. Nintendo Wii game console); the only certain means to prevent them from consuming electricity is to disconnect their power supply.

- Cold appliances: Refer to electric appliances such as fridges and freezers which are continuously in use but do not consume constant amount of electricity; instead their electricity consumption cycles between zero and a preset level.

- Active appliances: Refer to electrical appliances such as lights, kettles and electrical cookers which are actively switched on or off by users and are clearly either in use or not in use; they do not have a standby mode and when switched off they do not consume electricity at all.

The electric equipment and appliances in office buildings (computing equipment, lights and security devices) have the same patterns of electricity consumption as home electric appliances. Drawing on the idea of Firth et al [12], we see the electricity consumed by continuous appliances (e.g. security cameras, information displays and computer servers) and cold appliances (e.g. refrigerators) as base consumption, because these kinds of electric equipment and appliances (we term them *base appliances*) have to be switched on all the time; and we see the electricity consumed by active appliances (e.g. lights) and standby appliances (e.g. desktop computers and printers) as flexible consumption, because these kinds of electric equipment and appliances (we term them *flexible appliances*) can be switched on/off at any time, depending on the behaviour of users. Thus, the total electricity consumption of an office building in a certain period of time can be formulated as:

$$C_{\text{total}} = C_{\text{base}} + C_{\text{flexible}} \quad (1)$$

Where $C_{\text{base}}$ is the base electricity consumption and relates to the number and types of continuous and cold appliances the office has; $C_{\text{flexible}}$ is the flexible electricity consumption and relates to the number and types of active and standby appliances in the office.

We see the electricity users as agents. Considering the individual active and standby appliances and the behaviour of their electricity user agents, we can further break down the flexible consumption:

$$C_{\text{flexible}} = \beta_1 * C_{\text{f1}} + \beta_2 * C_{\text{f2}} + \beta_3 * C_{\text{f3}} + \cdots + \beta_n * C_{\text{fn}} \quad (2)$$

Where $C_{\text{f1}}, C_{\text{f2}}, C_{\text{f3}}, C_{\text{fn}}$ are the maximum electricity consumption of each flexible appliance; $n$ is the number of flexible appliances; and $\beta_1, \beta_2, \beta_3, \beta_n$ are the parameters reflecting the behaviour of the electricity user agents. We range **β** from 0 to 1. If **β** is near 0, it means that the electricity user agent of the flexible appliance always switches it off. If **β** is close to 1, it means that the electricity user agent of the flexible appliance always leaves it on.

Combining equation 1 and 2, we can derive an equation to explain the electricity consumption in an office in a certain period of time:

$$C_{\text{total}} = C_{\text{base}} + (\beta_1 * C_{\text{f1}} + \beta_2 * C_{\text{f2}} + \beta_3 * C_{\text{f3}} + \cdots + \beta_n * C_{\text{fn}}) \quad (3)$$

Equation 3 explains how the behaviour of electricity user agents can contribute to the overall electricity consumption in an office. It is the rationale for developing an agent-based model of office building electricity consumption.

## 3. Agent-Based Simulation of Office Building Electricity Consumption: A Case Study

As electricity users' behaviour is significantly influenced by the electricity management technologies and electricity management policies/regulations in an office building, the above equation potentially integrates the four elements we mentioned before (i.e. energy management policies/regulations, energy management technologies, electric equipment/appliances and user behaviour) and provides a theoretical basis for developing an agent-based simulation model of office building electricity consumption. Here we develop an agent-based model of office building electricity consumption based on an academic building in the School of Computer Science, located in Jubilee Campus, the University of Nottingham. We chose this case because it is very convenient for us to carry out surveys to understand users' behaviour in the school, and also the Estate Office, who is responsible for the energy management in the University of Nottingham, kindly provided us with data about electricity management technologies and really electricity consumption in the school building.

### 3.1 Electricity Consumption in the School Building

The School of Computer Science Building is situated in Jubilee Campus which was opened in 1999. Built on a previously industrial site, Jubilee Campus is an exemplar of brownfield regeneration and has impeccable green credentials. In terms of energy technologies, one important feature of the campus is the series of lakes which not only is the home of a variety of wildlife, but also provide storm water attenuation and cooling for the buildings in summers. Less visible, but equally important energy technologies are (1) the roofs of the buildings covered by low-growing alpine plants which help insulate and maintain steady temperatures within the buildings throughout the year, (2) super-efficient mechanical ventilation systems, (3) lighting sensors to reduce electricity consumption, and (4) photovoltaic cells integrated into the atrium roofs. Jubilee Campus has received many awards for its environment-friendly nature and energy efficiency of its buildings.

The School of Computer Science Building occupies a central position in Jubilee Campus, and is the academic home of some hundreds of staff and students. The base electricity consumption in the school building includes security devices, information displays, computer servers, shared printers and ventilation systems. The flexible electricity consumption includes lights and office computers. In terms of electricity management technologies, the school building is equipped with light sensors and half-hourly metering systems. Based on these electricity management technologies, the energy management division (the Estate Office) has made automated lights electricity management strategy (lights are automatically switched on when staff enter a room and switched off in 20 minutes after staff leave the room), and also select two environmental champions to promote energy saving awareness in the school.

Currently there are two practical electricity management issues arising in the school. One issue is the debate over the automated lighting management strategy. On the one hand, many technical people from the Estate Office believe that automated lighting management strategy is more energy-efficient than manual lighting management strategy (i.e. installing light switches to enable staff to control the lights). On the other hand, many electricity users in the School of Computer Science believe that if they could control the light manually, the electricity consumption in the school would be less, as under the automated lighting management strategy the lights are off only after 20 minutes of their leave, which causes unnecessary consumption of electricity. The other issue is measuring the proportions of electricity consumed by lights and computers. Although currently the school is equipped with advanced half-hourly electricity meters which can tell us how much electricity is consumed by the whole school, they are not able to tell us how much electricity is consumed by computers and how much electricity is consumed by lights. Technically speaking, the amount of electricity consumed by lights and computers is related to the behaviour of their users. Thus we can hardly to measure it in a simply way. Focusing on this two practical electricity management issues, in this case study we are targeting two research questions: (1) Is automated lighting strategy always energy-efficient than staff-controlled lighting management strategy? (2) What are the proportions of electricity consumed by lights and computers respectively?

**3.2 Agent Based Model of Office Building Electricity Consumption**

For our case study we have chosen the first floor of the School of Computer Science. This floor is populated by academics, research staff, research students, and admin staff. The building plan of the floor is shown in Figure 2. The details of the rooms and electric equipment and appliances on the first floor are listed in Table 1.

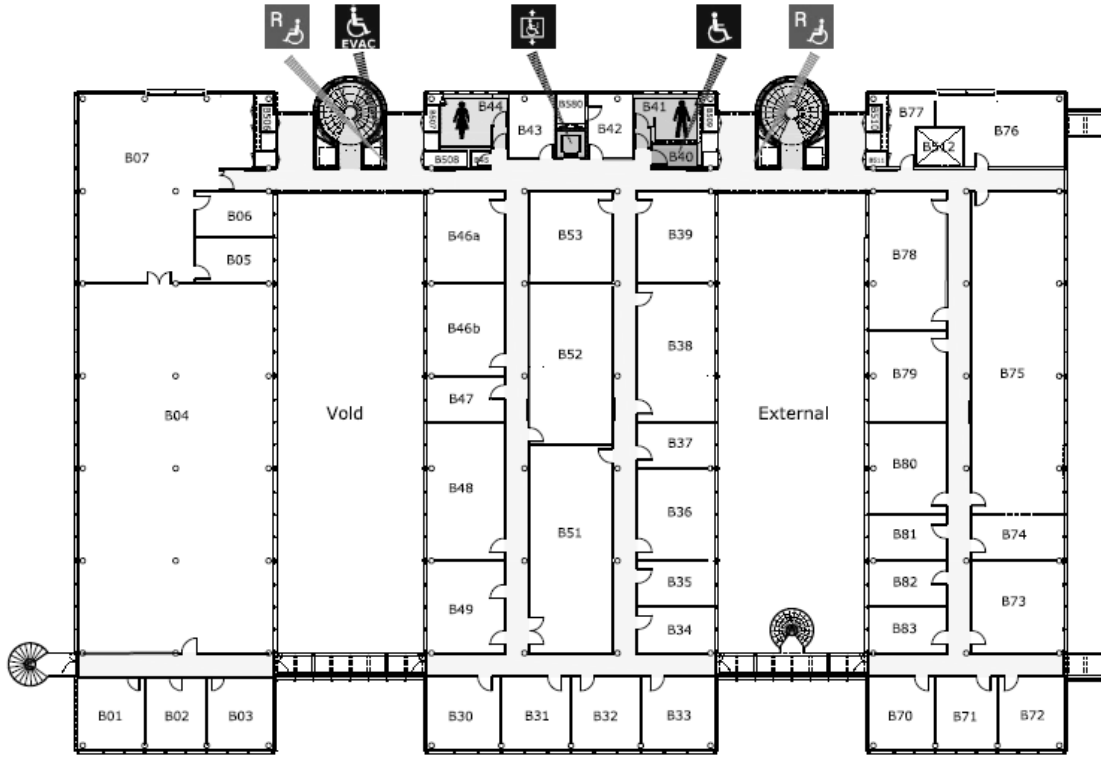

Figure 2: The Building Plan

Table 1: Details of Rooms and Electric Equipment and Appliances on the First Floor

| Item | Number |
| --- | --- |
| Rooms | 47 |
| Lights | 239 |
| Computers | 180 |
| Printers | 24 |
| Information Displays | 2 |
| Maximum Number of Energy Users | 213 |

We designed the model environment based on the office plan on the first floor of the school building (Figure 2), and implemented the model in AnyLogic 6.5.0. The base electricity consumption of the school building is fixed (and therefore we don't need to consider it in our simulation model; it will simply be added when we do the output analysis), and the flexible consumption of the school building is caused by the interactions between flexible appliances (mainly lights and computers) and the electricity users. We therefore focus on the flexible consumption, and design three types of agents: electricity user agents, computer agents and light agents. These agents have been assigned to different rooms based on the office plan.

**3.2.1 Behaviour of Electricity User Agents**

In order to understand the electricity consumption behaviour of the electricity users, we have carried out an extensive school wide empirical survey (questionnaire and observation). We deployed an online questionnaire, and emailed 200 staff and PhD students. In total we have received 143 valid responses (response rate = 71.5%). Our survey focuses on the electricity use behaviour of the electricity users (i.e. staff and PhD students) when they are in the School of Computer Science for work. A descriptive statistics of their behaviour patterns is shown in Table 2. Our observation shows that during each working day, electricity users gradually enter the school building, walk through the corridors, and enter different offices for work. Their behaviour in different stages can trigger the electricity consumption of different electric appliances. Based on this observation, we develop an electricity user state chart that allows us to represent the behaviour of electricity users, as shown in Figure 3.

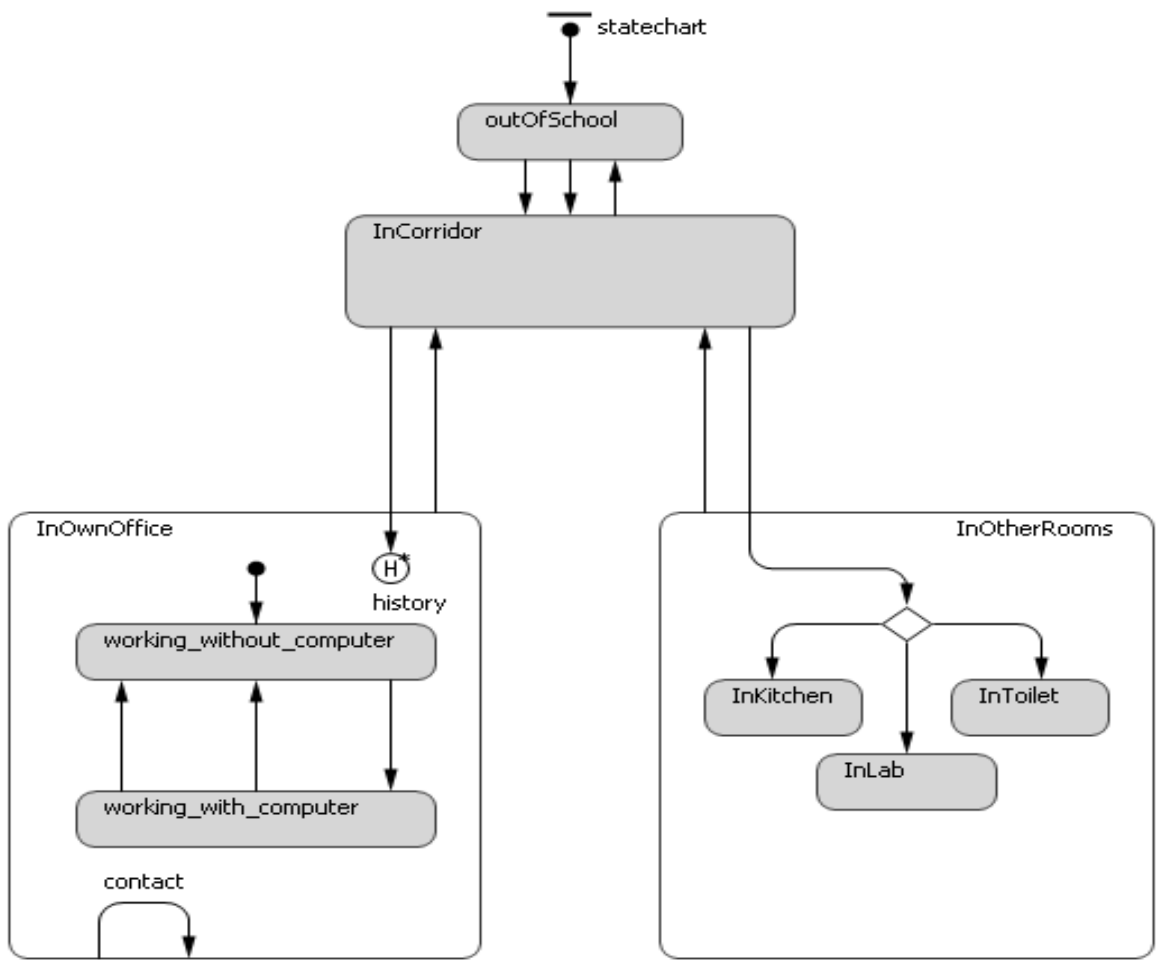

Figure 3: State Chart of Energy User Agents

Table 2: Statistics of Electricity User Behaviour Patterns

| Behaviour | Pattern | Frequency | Remarks |
|---|---|---|---|
| Time of Arrival at the School | Arriving at the school between 5am and 9am | 8% | These are designed as early bird electricity users agents |
| | Arriving at the school between 9am and 10am | 53% | These are designed as timetable complier electricity users agents |
| | Arriving at the school between 10am and 1pm | 39% | These are designed as flexible worker electricity users agents |
| Frequency of switching off computers when leaving the office | Every time | 1% | These are designed as environment champion electricity users agents |
| | Most of the time | 8% | These are designed as energy saver electricity users agents |
| | Not very often | 31% | These are designed as regular user electricity users agents |
| | Never | 60% | These are designed as big user electricity users agents |
| Talking to others about energy issues in the school | Very often | 8% | These are used for designing the contact frequency between electricity user agents |
| | Sometimes | 31% | |
| | Only occasionally | 45% | |
| | Never | 15% | |
| Using the kitchen | Almost every day | 36% | These are used for designing the frequency of using the kitchen in the simulation model |
| | Some times | 14% | |
| | Occasionally | 27% | |
| | Never | 23% | |
| What can help the school reduce energy consumption? | The energy management in the school is fully automated | 32% | These are used to support the simulation experiments |
| | The energy management is the school is controlled by users | 35% | |
| | Using more advanced energy management technologies | 19% | |
| | Giving energy users incentives for saving energy | 10% | |
| | Others | 4% | |

We consider four different states of an electricity user agent's behaviour in the model: out of school (*outOfSchool*), in corridor (*InCorridor*), in its own office (*InOwnOffice*) and in other rooms (*InOtherRooms*). In the *outOfSchool* state, the electricity user agent is not at work, thus does not trigger any electricity consumption. In the *InCorridor* state, as there are many lights in the corridors, the electricity user agent's presence in the corridor triggers the lights on, which causes electricity consumption. In the *InOwnOffice* state, the electricity user agent's presence in its own office triggers the office lights on, and its behaviour of using the computer in the office enables the computer in one of the following three modes: on, standby and off. Analogously, in the *InOtherRooms* state the electricity user agent's presence in other rooms (e.g. toilet, kitchen and lab) triggers the electricity consumption of lights and computers (if any) in these rooms.

The transitions between the *outOfSchool* state and the *InCorridor* state (both directions) is based on working timetable. Based on our empirical survey on working time, we have developed three stereotypes of electricity user agents: early birds, timetable compliers, and flexible workers. Early birds (mainly cleaners, security staff and some hard working students and staff) often come to the school between 5 am and 9 am, and leave the school according to their regular working time. Timetable compliers (mainly administrative staff and academic staff) often come to the school between 9 am and 10 am, and leave the school building often at 5:30 pm. Flexible workers (mainly academic staff, research staff and PhD students) come to school at any time between 10 am and 1 pm, and leave the school at any time after their arrival. Based on our statistics from our survey, we assign relevant parameters for the electricity user agents (Table 3). We also consider that each electricity user agent has a very small probability (p=0.02) to work on Saturdays and Sundays. This consideration reflects the reality that a small number of hard working PhD students and research staff come to school on Saturdays and Sundays.

Table 3: Stereotypes of Electricity User Agents and Parameters (I)

| Agent Stereotype | Percentage | Arrival Time | Leave Time |
| --- | --- | --- | --- |
| Early Birds | 8% | Monday to Friday, between 5am and 9am, random uniform distribution | Monday to Friday, between 5pm and 6pm, random uniform distribution |
| Timetable Compliers | 53% | Monday to Friday, between 9 am and 10 am, random uniform distribution | Monday to Friday, between 5pm and 6pm, random uniform distribution |
| Flexible Workers | 39% | Monday to Friday, between 10 am and 1 pm, random uniform distribution | Monday to Friday, between arrival time and 23pm, random uniform distribution |

The transition from the *InCorridor* state to the *InOwnOffice* state is an electricity user agent's behaviour of entering its own offices. In the simulation model we set the transition rule timeout = 2, which reflects the reality that normally after about a two-minute walk in the corridors, an electricity user can reach his/her own office. In the *InOwnOffice* state, the electricity user agent's presence triggers the lights in its own office on. The electricity user agent can either work with a computer (the sub-state *working_with_computer*), or work without a computer (the sub-state *working_without_computer*). The transition from the

working_without_computer sub-state to the working_with_computer sub-state is the electricity user agent's behaviour of switching on the computer. We set the transition rule timeout = 2. This design is based on our empirical observation that normally an electricity user switches on his/her computer within 2 minutes after he/she enters his/her office. The transitions from the working_with_computer sub-state to the working_without_computer sub-state is the electricity user agent's behaviour of switching off or setting the computer on standby. For setting the computer on standby, the transition rule is a probability (p = 0.05) derived from our empirical survey. For switching off the computer, the transition rule is a threshold control. We assume that each electricity user agent has a personality parameter *energySavingAwareness*, ranging from 0 to 100, to represent its awareness on energy saving. If an electricity user agent's *energySavingAwareness* is greater than a threshold, it has a large probability to switch off the computer when it does not need to use the computer. In the simulation, the threshold is adjustable, with value ranging from 0 to 100. Based on our empirical survey on staff's energy awareness, we create four stereotypes of electricity user agent for the simulation model: Environmental Champion, Energy Saver, Regular User, and Big User. Different stereotypes of electricity user agents have different levels of *energySavingAwareness*, and the probabilities for them to switch off the electric appliances that they do not have to use are different, as shown in Table 4. In the *InOwnOffice* state, an electricity user agent can also interact with other electricity user agents. Our empirical survey shows that in terms of energy issues in the school, the most widely used interacting means is using emails in offices. We thus use an internal transition "contact" within the *InOwnOffice* state to reflect the interactions between electricity user agents. The larger the *energySavingAwareness* an electricity user agent has, the larger probability it will send emails about energy saving issues to other electricity user agents who have interactions with it. We set the interacting social network type as "small world". Based on the statistics from our empirical survey, we assign relevant parameters for these stereotypes of electricity user agents (Table 4).

Table 4: Stereotypes of Energy User Agents and Parameters (II)

| Stereotype of Agent | Percentage | energySavingAwareness | Probability of Switching Off Unnecessary Electric Appliances | Probability of Sending Email about Energy Issues to Others |
|---|---|---|---|---|
| Environment Champion | 1% | Between 95 and 100, random uniform distribution | 0.95 | 0.9 |
| Energy Saver | 8% | Between 70 and 94, random uniform distribution | 0.7 | 0.6 |
| Regular User | 31% | Between 30 and 69, random uniform distribution | 0.4 | 0.2 |
| Big User | 60% | Between 0 and 29, random uniform distribution | 0.2 | 0.05 |

The transition from the *InOwnOffice* state to the *InCorridor* state reflects an electricity user agent's behaviour of leaving its own office. For an electricity user agent, this can happen at any time between its arrival time and leave time. Thus the transition rule is a stochastic event and the probability for it to happen is determined by its arrival time and leave time. Here we consider two kinds of leaves: temporary leave and long time leave. Temporary leave means that the electricity user agent leaves its own office for less than 20 minutes, while long time leave means that it leaves its own office for more than 20 minutes. According to our empirical observation, people who temporally leave their offices do not usually switch off electric appliances, but people who leave their offices for a relatively long time do.

The transition from the *InCorridor* state to the *InOtherRooms* state reflects the behaviour of using other facilities such as kitchens, toilets and labs in the school. For an electricity user agent, this behaviour can also happen at any time between its arrival time and leave time. Again we consider the transition rule as a stochastic event and the probability for it to happen is determined by the agent's arrival time leave time.

The transition from the *InOtherRoom* state to the *InCorridor* state reflects an electricity user agent's behaviour of stopping using other facilities and leaving the facility rooms. The transition rule is a timeout. Here we set the time range from 1 to 10 (random uniform distribution), which reflect the reality that a user usually finishes using facilities such as toilets or kitchens within 10 minutes.

The state chart, which we have developed based on our empirical survey, reflects the all behaviour of a real electricity user with regards to electricity consumption when he/she works in the school building.

**3.2.2 Behaviour of Light Agents**

In the simulation model, we treat the lights in the school building as passive agents, which means that these agents do not have proactive behaviour. Instead, their behaviour is passive reacting to the behaviour of electricity user agents. Lights' behaviour pattern is relatively simple, as they can only be either off or on, as shown in the light state chart (Figure 4)

For a light, the transition from the *off* state to the *on* state is associated with the presence of electricity user agents if the light is automated by light sensors, or with the behaviour of switching on if there is a light switch which enables electricity user agents to have control over the light. Conversely, the transition from the *on* state to the *off* state is associated with electricity user agents' leaving if the light is automated by light sensors, or with the behaviour of switching off if there is a light switch. According to the data provided by the Estate Office, when a light is on, its power is 60 Watts; when it is off, its power is 0.

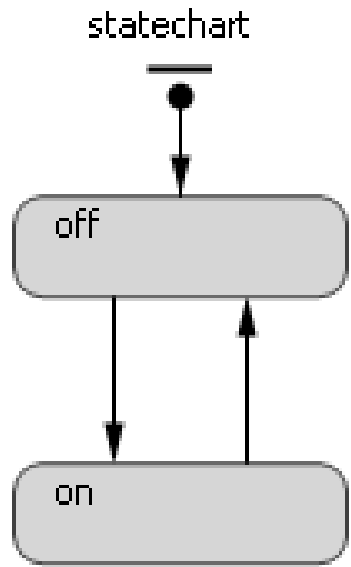

Figure 4: State Chart of Light Agents

### 3.2.3 Behaviour of Computer Agents

Similar to lights, computers are also treated as passive agents in our simulation model. Computers can be on, off, or in standby. Thus in the state chart of a computer, we consider these three states, as shown in Figure 5.

For a computer, the transitions between *off*, *on* and *stand_by* states are related to electricity user agents' behaviour of using the computer. According to our survey, when a computer is off, its overall power is 0; when it is on standby mode, its overall power is 25 Watts; and when it is on, its overall power is about 400 Watts.

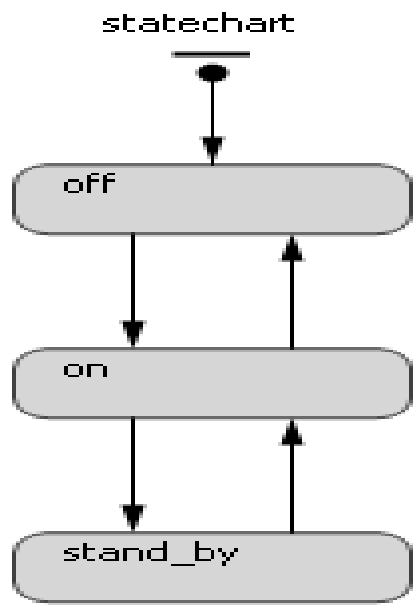

Figure 5: State Chart of Computer Agents

### 3.2.4 Model Implementation

The model has been implemented in the simulation package AnyLogic 6.5.0 [13] on a standard PC with Windows XP SP3. We set each time step in the simulation model as one minute, and simulate the daily work of staff in the School of Computer Science and observe and analyse how their behaviour can result in a system level electricity consumption of the whole floor. The light agents and computer agents are assigned to each room, based on their real physical distribution in the school. The electricity user agents come to the school every morning, walk through the corridors and enter their own offices for work. They may also leave their offices, walk through the corridors and enter other rooms for using facilities such as toilet and kitchens. They interact with each other in terms of energy issues in the school, and this kind of interaction can increase their *energySavingAwareness*. They also interact with passive light and computer agents, and this kind of interaction can directly result in the system level electricity consumption of the school. An overview of the model is shown in Figure 6. We also animate the electricity user agents, and the interface of the model is shown in Figure 7.

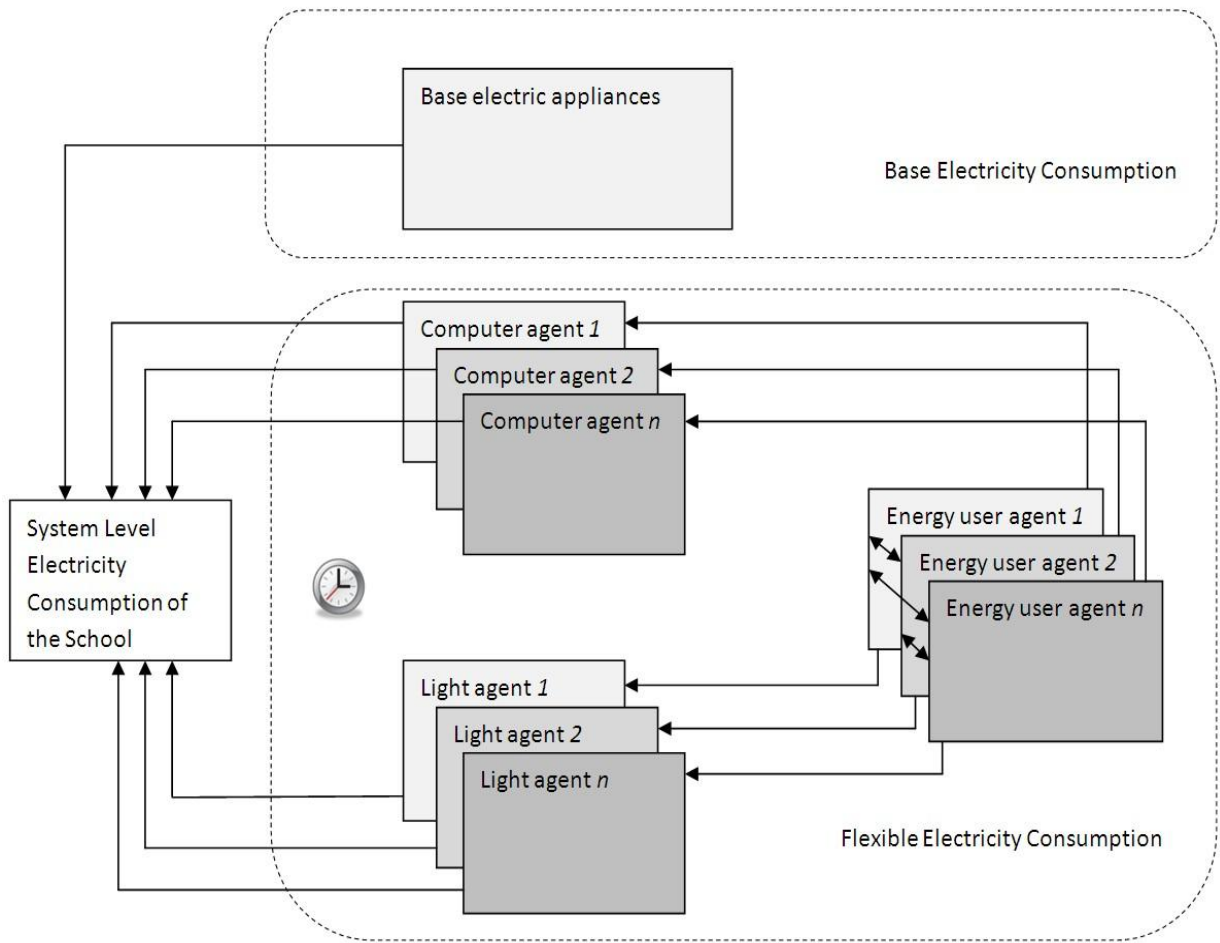

Figure 6: Overview of the Model

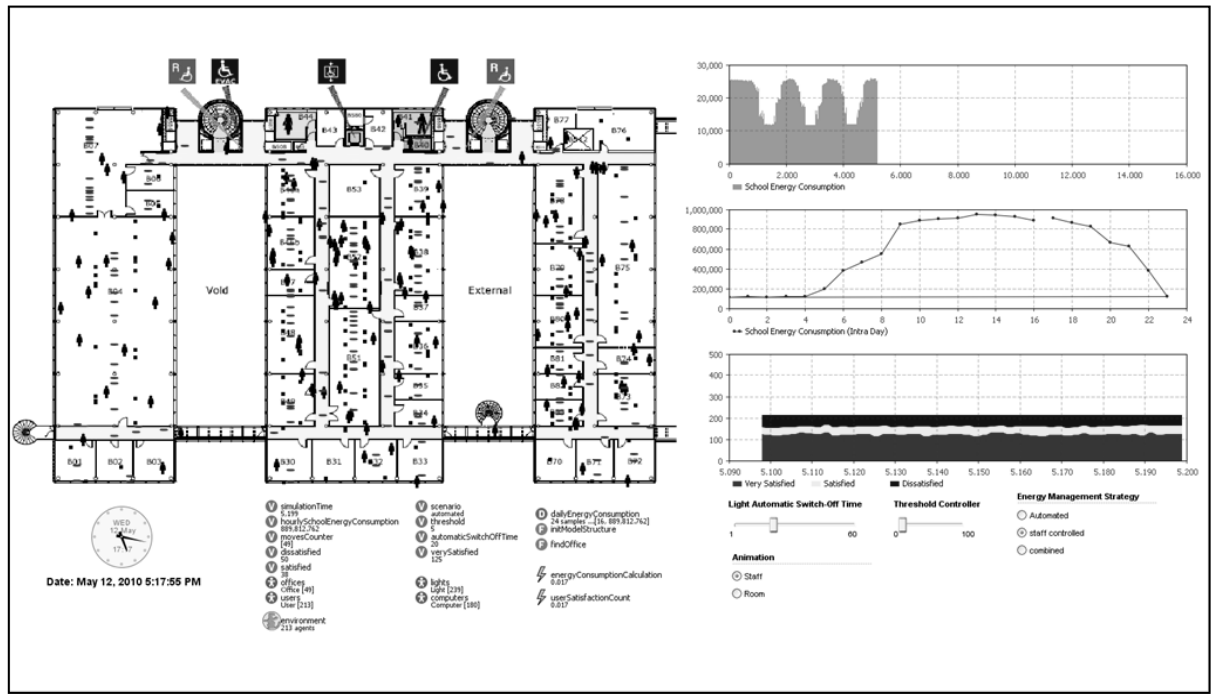

Figure 7: Interface of the Model

## 4. Simulation Experiments

With our model, we have carried out three sets of experiments. We use these sets of experiments to test the validity of the model, design electricity management strategies for the Estate Office and help the Estate Office gain insights about the electricity consumption in the school.

**Experiment 1: Reproduce the current electricity management strategy of the school**

Currently, the computer science school is equipped with light sensors which automatically switch on the lights when they detect the presence of staff, and switch off the lights when they detect the absence of staff for 20 minutes. Thus based on the light sensor technology, the Estate Office has adopted an *automated* electricity management strategy in the school of computer science. In that sense, staff do not have control over the switch-on/off of the lights, and they only have control over the switch-on/off the computers. Our first set of experiments focuses on this and aims to reproduce the electricity management strategy. We set the model to the "automated" scenario, run the model and plot the system level school electricity consumption in Figure 8, from which we can see that the pattern of simulated school electricity consumption is quite similar to the real school electricity consumption provided by the Estate Office (Figure 8). The similarity from the comparison signifies that we have successfully reproduced the current electricity management strategy in the school of computer science, and also validates our model.

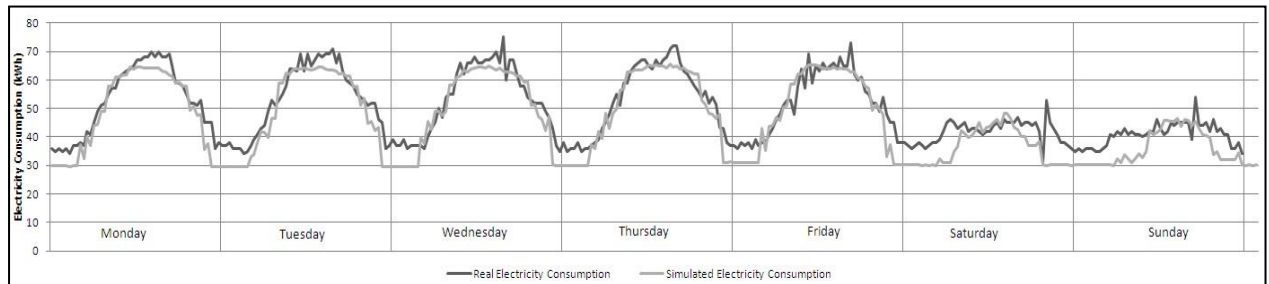

Figure 8: Comparison of Simulation Results and Empirical Results (Experiment 1)

Note: In this figure the simulation result is the average of results of 20 replications with different random seeds.

**Experiment 2: Automated Strategy vs. Staff-Controlled Strategy**

Automated and manual lighting management each have advantages in office buildings [14]. Although the office buildings that have achieved the lowest reported lighting electricity consumption have done so with manual electricity management [15, 16], some studies show that under manual switching lighting management it was quite common for users to switch on lights even when there was adequate glarefree daylight [17]; once switched on, the lights were seldom switched off, regardless of the illumination provided by the daylight [16]. Thus there is an argument for an automated electricity management strategy which maintains efficient electricity consumption to satisfactorily meet electricity users' electricity needs and meanwhile minimizes the cost of electricity without any user intervention in office buildings. Bourgeois et al. [18] show that under automated lighting management the electricity consumption is much less than that in automated lighting management. In the School of Computer Science, there seems to be a debate between the automated strategy and staff-controlled strategy. On the one hand, anecdotally, many people, particularly technical people from the Estate Office, strongly believe that automated lighting is more energy-efficient than staff-controlled lighting. On the other hand, our empirical survey shows that some people in the School of Computer Science believe that if they could control the light manually, the electricity consumption in the school would be less, as under the automated lighting strategy the lights are off only after 20 minutes of their leave, which causes unnecessary consumption of electricity. Our second set of experiments focuses on this debate: we compare the simulation results under the two different lighting management strategies. The rationale for the two strategy scenarios is as follows: in the automated lighting management strategy scenarios, lights in an office are off 20 minutes after the last occupying electricity user agent' leave, while in the staff-controlled lighting management strategy, lights in an office are switched off by the last occupying electricity user agent based on a probability. The probability is determined by the *energySavingAwareness* of the last occupying electricity user agent. The larger the *energySavingAwareness* the last occupying electricity user agent has, the larger the probability it will switch off the lights. The probabilities are assigned based on

Table 4. This design reflects the reality that under the staff-controlled lighting management strategy, staff can switch off the lights when they leave their offices. The more the staff are concerned about energy saving, the more possible they will switch off the lights when they leave their offices. The comparison of the simulation results of the two scenarios are shown in Figure 9, from which we can see that although the peak time electricity consumption is almost the same, the electricity consumption in staff-controlled lighting management strategy scenario is substantially higher than that in automated lighting management strategy scenario. Thus, one electricity management strategy implication we can draw from the simulation is that in the current circumstance, the automated lighting management strategy is more energy-efficient than staff-controlled electricity management strategy.

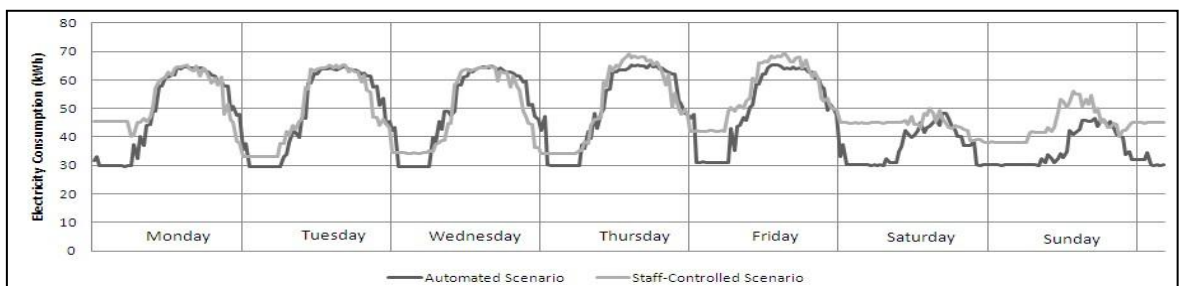

Figure 9: Simulation Results (Experiment 2A)

Note: In this figure the simulation result is the average of results of 20 replications with different random seeds.

We note that the probabilities for staff to switch off lights when they leave their offices are related to their *energySavingAwareness.* One question to which we would like to seek answer from the model is that: if we increase the electricity user agents' *energySavingAwareness* by enhance the interactions about energy issues between electricity user agents, is automated lighting management strategy still more energy-efficient than staff-controlled lighting management strategy? We increase the contact rate (i.e. the frequency of contact in a certain simulation period), run the model and gain the simulation results in Figure 10, which shows a negative answer to that question: when enhancing the interactions about energy saving between electricity user agents, automated strategy is less effective than staff-controlled strategy. Increasing electricity user agents' *energySavingAwareness* through social interactions can significantly reduce the overall electricity consumption of the school. The senior management of the university has already realized the importance of increasing staff's energy saving awareness, thus a university-wide campaign called "gogreener" has been carried out and two environmental champions have been appointed in each school to monitor the energy consumption of the school and enhance the interactions of staff in terms of energy issues.

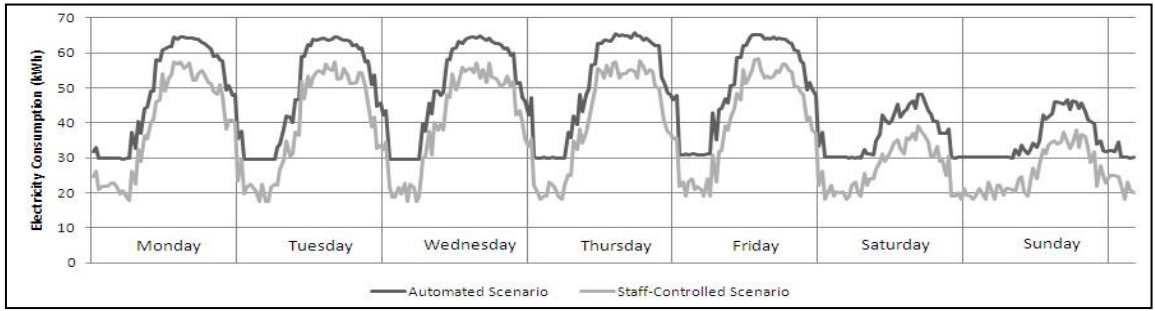

Figure 10: Simulation Results (Experiment 2B)

Note: This figure shows the simulated electricity consumption in both automated lighting management scenario and staff-controlled lighting management scenario when the interactions between energy user agents have been enhanced. In this figure the simulation results are the average of results of 20 replications with different random seeds.

**Experiment 3: Understanding the proportions of electricity consumed by lights and computers**

The Estate Office has installed some half-hourly electricity meters in the school building to monitor the electricity consumption in the School of Computer Science. Although these meters can tell us how much electricity is consumed by the school, they are not able to tell us how much electricity is consumed by computers and how much electricity is consumed by lights, which is also a question the Estate Office keen to know. As indicated in the model, the amount of electricity consumed by lights and computers is related to behaviour of electricity user agents, which makes it hard to be measured in a simply way. With the help of the simulation model, we can gain some insights into this issue. We run the model in *automated* scenario (i.e. the current lighting management strategy of the school), and plot the electricity consumption by both lights and computers in Figure 11, from which we can see the proportions of electricity consumed by computers and lights vary over time. We also plot one week electricity consumption, as shown in Figure 11.

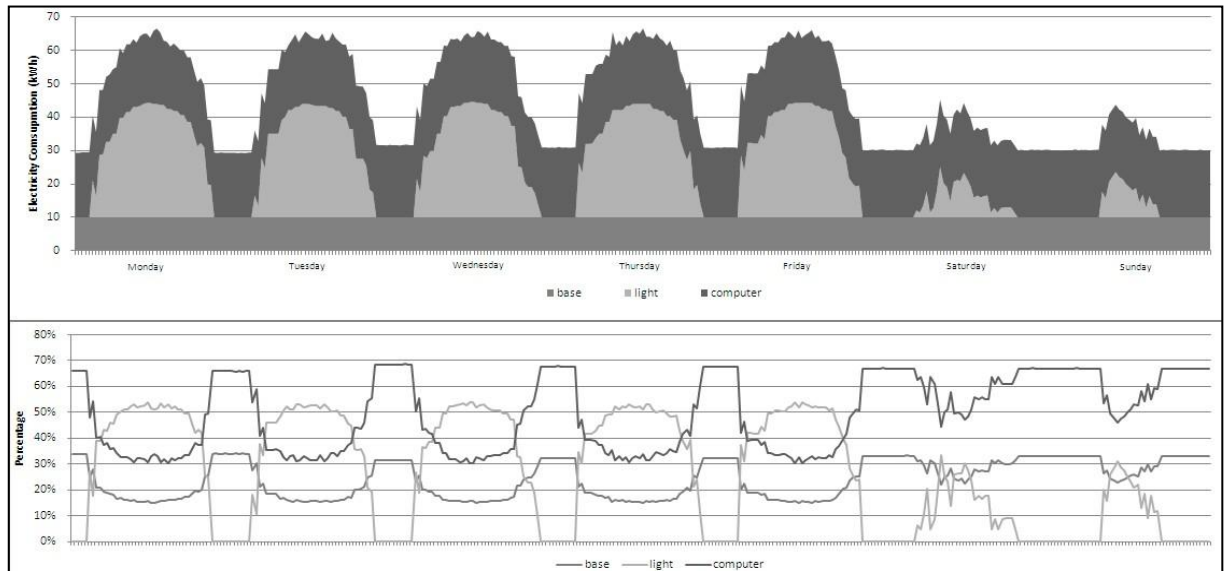

Figure 11: Electricity Consumed by Computers, Lights and Based Electric Appliances

Note: The figure on the top is the amount of electricity consumed by lights, computers and base electric appliances, while the figure on the bottom shows the percentages. From the simulation results we can see that in the evenings and weekends, most of the electricity is consumed by computers (65%); in the daytime (Monday to Friday), the electricity consumed by computers (about 33%) is much less than that consumed by lights (52%).

## 5. Discussion

**Theory Based Agents vs. Empirical Survey Based Agents**

In social simulation, many agent-based simulation studies develop the agents in their models based on well-established social theories. For example, in marketing researchers develop consumer agents based on social psychological theories such as social comparison, imitation [e.g. 19, 20, 21] and the theory of planned behaviour [e.g. 22, 23]. In energy economics, Bunn and Oliveira [24] developed electricity market agents (i.e. electricity generating companies and electricity suppliers) based on market bidding theory to simulate the New Electricity Trading Arrangements (NETA) of England and Wales. Clearly, the existence of these well-established social theories significantly facilitates the development of the social agents.

In this particular case of modelling office building electricity consumption, however, there are no well-established theories that could be used to model staff behaviour towards electricity use. We thus have to conduct time-consuming and costly empirical surveys and observations on the behaviour of real world objects (i.e. staff and PhD students in the school) and develop agents based on our empirical survey. Our survey covered most of the aspects of staff's electricity use in the school. As a result, the state chart we have developed to represent the behaviour of electricity users in our model has a very strong empirical basis. That is also the reason why our simulation results are quite similar to the real world observation. Empirical survey based agents are increasingly used in social simulation, particularly in operations management [e.g. 25, 26]. Compared to traditional theory based agents, empirical survey based agents, although require much more time and work for the development, are easier to be calibrated and validated.

**Limitations**

Although we have developed a state chart to represent the behaviour of electricity user agents based on a comprehensive empirical survey, clearly artificial agents cannot perfectly replicate the real-life of electricity users in the school. Therefore, it is important to acknowledge the limitations of the model. Firstly, an electricity user agent's stereotype in the model is fixed. In other words, in the simulation there is no way for an electricity user agent to switch its stereotype (e.g. from an early bird to a flexible worker). But we note that in the real world the switch of stereotypes can happen, although the probability for its happening is low. A second limitation is our assumption that enhancing interactions about energy issues between staff can increase staff's energy saving awareness. This assumption is true in the situation where electricity users have to bear the cost of electricity, as some research on energy efficiency in domestic sector has already proved it [e.g. 27]. However, while working in office buildings staff do not have to bear the cost of electricity, which may result in the assumption in question. Currently we have not found any sound evidence to support this assumption.

**Further Research**

The agent-based model of office building electricity consumption described in this paper has potential for further development. Theoretically, we can incorporate more flexible electric appliances and more complex human-electric appliance interactions into the model, which will make the model more applicable. It can then be applied to modelling a large organisation which has very complex behaviour of consuming energy, e.g. large number of staff and complex energy management strategies/regulations. This type of simulation models can potentially be developed as a building energy simulation software package which provides human-centric organisations with organisational energy policy making support. Moreover, we can add more psychological factors into the energy user agents, and study how to optimize energy consumption for an organisation while maintain its staff's satisfaction about energy use.

## 6. Conclusions

This paper has described an agent-based model of office building electricity consumption. We began the paper with an argument for an integration of the four elements involved in office building energy consumption, and then described the agent-based simulation method and the rationale for developing an agent-based model for studying office building electricity consumption. We then developed an agent-based model of office building electricity consumption based on the case of the School of Computer Science, in Jubilee Campus, the University of Nottingham, and presented the simulation results. Along the way, we focused on two objectives. One is the integration of the four elements involved in office building energy consumption in to one model. The other one is developing a multi-agent framework to study practical energy management issues for an organisation. From the research we reported in this paper, we conclude that, although it is not possible to perfectly replicate the real organisation, agent-based simulation as a novel approach which integrates the four elements involved in office building energy consumption, is a very useful tool for office building energy management.

# References

[1] H.B. Rijal, P. Tuohy, M.A. Humphreys, J.F. Nicol, A. Samuel, J. Clarke, Using results from field surveys to predict the effect of open windows on thermal comfort and energy use in buildings, Energy and Buildings 39 (7) (2007) 823-836.

[2] A. Mahdavi, L. Lambeva, A. Mohammadi, E. Kabir, C. Proglhof, Two case studies on user interactions with buildings' environmental systems, Bauphysik 29 (1) (2007) 72-75.

[3] J.F. Nicol, Characterising occupant behaviour in buildings: towards a stochastic model of occupant use of windows, lights, blinds, heaters and fans, in: Proceedings of Building Simulation '01, Rio de Janeiro, Brazil, (2001), pp. 1073-1078.

[4] C.F. Reinhart, Lightswitch-2002: a model for manual and automated control of electric lighting and blinds, Solar Energy 77 (1) (2004) 15-28.

[5] P. Hoes, J. L. M. Hensen, M. G. L. C. Loomans, B. de Vries, and D. Bourgeois, User Behaviour in Whole Building Simulation. Energy and Buildings, Elsevier 41(2009):295-302.

[6] E. Azar, C. Menassa, A Conceptual Framework to Energy Estimation in Buildings Using Agent Based Modelling, in Proceedings of the 2010 Winter Simulation Conference, Baltimore, Maryland, (2010). 3145-3155.

[7] V.L. Erickson, Y. Lin, A. Kamthe, R. Brahme, A. Surana, A.E. Cerpa, M.D. Sohn and S. Narayanan, Energy Efficient Building Environment Control Strategies Using Real-time Occupancy Measurements, Proceedings of ACM BuildSys 2009, First ACM Workshop on Embedded Sensing Systems for Energy-Efficiency in Buildings, Berkeley, Calif., Nov., 2009.

[8] S.J. Russell and P. Norvig, Artificial Intelligence: A Modern Approach (2nd ed.), Upper Saddle River, New Jersey: Prentice Hall, 2003.

[9] R.L. Axtell, Why Agents? On the Varied Motivations for Agent Computing in the Social Sciences. In Macal C M and Sallach D (Eds.) Proceedings of the Workshop on Agent Simulation: Applications, Models, and Tools: 3-24. Argonne, IL: Argonne National Laboratory, 2000.

[10] J.M. Galán, L.R. Izquierdo, S.S. Izquierdo, J.I. Santos, R. del Olmo, A. López-Paredes and B. Edmonds, Errors and Artefacts in Agent-Based Modelling, Journal of Artificial Societies and Social Simulation 12(2009)1 <http://jasss.soc.surrey.ac.uk/12/1/1.html>.

[11] J.D. Farmer and D. Foley, The economy needs agent-based modelling. Nature, 406 (2009) 685-686.

[12] S. Firth, K. Lomas, W. Wright and R. Wall, Identifying trends in the use of domestic appliances from household electricity consumption measurements. Energy and Buildings, 40(2008) 926-936.

[13] XJ Technologies. Official AnyLogic website. http://www.xjtek .com/. 2010. Data of Access: 4 Sep 2010.

[14] J.A. Love, Manual switching patterns in private offices. Lighting Research and Technology, 30(1998) 45-50.

[15] R.W. John and A.C. Salvidge, The BRE low-energy office—five years on. Building Service Engineering Research and Technology 7(1986) 121-128.

[16] B, Anderson, M. Adegran, T. Webster, W. Place, R. Kammerud, and P. Albrand, Effects of daylighting options on the energy performance of two existing passive commercial buildings. Building and Environment 22(1987) 3-12.

[17] A.M. Galasiu, J.A. Love, and M. Navab, Design and performance of a daylighted high school. Proceedings of 1995 Illuminating Engineering Society of North America Annual Conference. New York: IESNA (1995) 617-634.

[18] D. Bourgeois, C. Reinhart and I. Macdonald, Adding advanced behavioural models in whole building energy simulation: A study on the total energy impact of manual and automated lighting control. Energy and Buildings 38(2006) 814-823.

[19] M. Janssen and W. Jager, An integrated approach to simulating behavioural processes: a case study of the lock-in of consumption patterns. Journal of Artificial Societies and Social Simulation, 2(1999).

[20] M. Janssen and W. Jager, Simulating diffusion of green products. Journal of Evolutionary Economics, 12(2002) 283-306.

[21] M. Janssen and W. Jager, Simulating market dynamics: interactions between consumer psychology and social networks. Artificial Life, 9(2003) 343-356.

[22] T. Zhang and D. Zhang, Agent-based simulation of consumer purchase decision-making and the decoy effect. Journal of Business Research, 60(2007) 912-922.

[23] T. Zhang and W.J. Nuttall, Evaluating Government's Policies on Promoting Smart Meting in Retail Electricity Markets via Agent-Based Simulation. Cambridge Working Paper in Economics, Cambridge: University of Cambridge, 2008.

[24] D. Bunn and F. Oliveira, Agent-Based Simulation: An Application to the New Electricity Trading Arrangements of England and Wales. IEEE Transactions on Evolutionary Computation, 5(2001) 493-503.

[25] P. Siebers, U. Aickelin, H. Celia and C. Clegg, Modelling and Simulating Retail Management Practices: A First Approach. International Journal of Simulation and Process Modelling, 5(2009) 215-232.

[26] P. Siebers, U. Aickelin, H. Celia, and C. Clegg, Simulating Customer Experience and Word-Of-Mouth in Retail – A Case Study. Simulation: Transactions of the Society for Modelling and Simulation International, 86(2010) 5-30.

[27] H. Wilhite, H. Nakagami, T. Masuda, and Y. Yamaga, A cross-cultural analysis of household energy use behaviour in Japan and Norway, Energy Policy, 24(1996) 795–803.